\documentclass[%
superscriptaddress,
amsmath,amssymb,
 aps,prl,
 twocolumn,
 tightenlines,
]{revtex4-1}

\usepackage{graphicx,url,amssymb,amsmath,longtable,rotating,color,units,wasysym,subfigure,epsfig,hyperref,epstopdf}
\usepackage[dvipsnames]{xcolor}
\usepackage{enumitem}

\usepackage{graphicx}
\usepackage{dcolumn}
\usepackage{bm}
\usepackage{url,hyperref}
\usepackage{array,multirow}
\usepackage{color} 
\usepackage{aas_macros} 
\usepackage{ifthen}
\usepackage{etoolbox}

\definecolor{darkgreen}{rgb}{0.0, 0.5, 0.0}

\newcommand{\Ogw}{\Omega_\text{GW}}
\def\Mc{\mathcal{M}_c}

\def\bnsrate{1540^{+3200}_{-1220} \text{ Gpc}^{-3}\text{ yr}^{-1}}

\def\bbhrate{103^{+110}_{-63} \text{ Gpc}^{-3}\text{ yr}^{-1}}

\def\OmegaBNSMean{0.7}
\newcommand{\OmegaRangeBNS}{\ensuremath{\OmegaBNSMean_{-0.6}^{+1.5} \times 10^{-9} }}

\def\OmegaBBHMean{1.1}
\newcommand{\OmegaRangeBBH}
{\ensuremath{\OmegaBBHMean_{-0.7}^{+1.2} \times 10^{-9} }}

\def\OmegaTotMean{1.8}
\newcommand{\OmegaRange}{\ensuremath{\OmegaTotMean_{-1.3}^{+2.7} \times 10^{-9} }}

\newcommand{\tauBNS}{\ensuremath{13_{-9}^{+49} }}
\newcommand{\tauBBH}{\ensuremath{223_{-115}^{+352} }}
\newcommand{\tauTotal}{\ensuremath{12_{-8}^{44} }}

\newcommand{\lambdaBNS}{\ensuremath{15 ^{+30} _{-12}}}
\newcommand{\lambdaBBH}{\ensuremath{0.06 ^{+0.06} _{-0.04}}}
\newcommand{\lambdaTot}{\ensuremath{15 ^{+31} _{-12}}}

\newcommand{\durBNS}{190}
\newcommand{\durBBH}{14}

\def\obs{40 months}
\def\obsopt{18 months}

\newtoggle{fullauthorlist}
\togglefalse{fullauthorlist}

\newtoggle{endauthorlist}
\toggletrue{endauthorlist}

\begin{document}

\preprint{APS/123-QED}

\title{GW170817: Implications for the Stochastic Gravitational-Wave Background \\ from Compact Binary Coalescences}

\date{\today}
             
\pacs{Valid PACS appear here}

\iftoggle{endauthorlist}{
  %
  %
  \let\mymaketitle\maketitle
  \let\myauthor\author
  \let\myaffiliation\affiliation
  \author{The LIGO Scientific Collaboration}
  \author{The Virgo Collaboration}
\noaffiliation
}{
  %
  %
  \iftoggle{fullauthorlist}{
  }{
    \author{The LIGO Scientific Collaboration}
    \affiliation{LSC}
    \author{The Virgo Collaboration}
    \affiliation{Virgo}
  }
}

\begin{abstract}
\iftoggle{endauthorlist}{
}{
  \newpage
}
The LIGO Scientific and Virgo Collaborations have announced the event GW170817, the first detection of gravitational waves from the coalescence of two neutron stars. 
The merger rate of binary neutron stars estimated from this event suggests that distant, unresolvable binary neutron stars create a significant astrophysical stochastic gravitational-wave background.
The binary neutron star component will add to the contribution from binary black holes, increasing the amplitude of the total astrophysical background relative to previous expectations.
In the Advanced LIGO-Virgo frequency band most sensitive to stochastic backgrounds (near 25 Hz), we predict a total astrophysical background with amplitude $\Ogw (f=25 \text{ Hz}) = \OmegaRange$ with 90\% confidence, compared with $\Ogw(f=25\text{ Hz})=\OmegaRangeBBH$ from binary black holes alone. 
Assuming the most probable rate for compact binary mergers, we find that the total background may be detectable with a signal-to-noise-ratio of 3 after 40 months of total observation time, based on the expected timeline for Advanced LIGO and Virgo to reach their design sensitivity.
\end{abstract}

\maketitle

{\em Introduction} ---
On 17 August 2017, the Laser Interferometer Gravitational-Wave Observatory (LIGO)~\cite{aLIGO_2015} Scientific and Virgo~\cite{aVirgo_2015} Collaborations detected a new gravitational-wave source: the coalescence of two neutron stars~\cite{gw170817}.
This event, GW170817, comes almost two years after GW150914, the first direct detection of gravitational waves from the merger of two black holes~\cite{gw150914}. 
In total, five high confidence detections and one sub-threshold candidate from binary black hole merger events have been reported: GW150914~\cite{gw150914}, GW151226~\cite{gw151226}, LVT151012~\cite{o1bbh}, GW170104~\cite{gw170104}, GW170608~\cite{gw170608}, and GW170814~\cite{gw170814}. 
The last of these events was reported as a
three-detector observation by the two Advanced LIGO detectors located in Hanford, WA and Livingston, LA in the United States and the Advanced Virgo detector, located in Cascina, Italy. 

In addition to loud and nearby events that are detectable as individual sources, there is also a population of unresolved events at greater distances.
The superposition of these sources will contribute an astrophysical stochastic background, which is discernible from detector noise by cross-correlating the data streams from two or more detectors~\cite{Allen_Romano_1999,christensen92}. 
For a survey of the expected signature of astrophysical backgrounds, see \cite{Regimbau_2011, ZhuEA_2011, Rosado_2011, MarassiEA_2011, WuEA_2012, ZhuEA_2013, Kowolska_2015}. See also \cite{RomanoCornish} for a recent review of data analysis tools for stochastic backgrounds. 

Following the first detection of gravitational waves from GW150914, the LIGO and Virgo Collaborations calculated the expected stochastic background from a binary black hole (BBH) population with similar masses to the components of GW150914~\cite{gw150914_stoch}. 
These calculations were updated at the end of the first Advanced LIGO observing run (O1)~\cite{stoch_O1} to take into account the two other black hole merger events observed during the run: GW151226, a high-confidence detection; and LVT151012, an event of lower significance. 
The results indicated that the background from black hole mergers would likely be detectable by Advanced LIGO and
Advanced Virgo after a few years of their operation at design sensitivity. 
In the most optimistic case, a detection is possible even before instrumental design sensitivity is reached. 

In this paper, we calculate the contribution to the background from a binary neutron star (BNS) population using the measured BNS merger rate derived from GW170817~\cite{gw170817}.
We also update the previous predictions of the stochastic background from BBH mergers taking into account the most recent published statements about the rate of events~\cite{gw170104}.
We find the contributions to the background from BBH and BNS mergers to be of similar magnitude.
As a consequence, an astrophysical background (including contributions from BNS and BBH mergers) may be detected earlier than previously anticipated.

{\em Background from compact binary mergers} ---
The energy-density spectrum of a gravitational wave background can be described by the dimensionless quantity 
\begin{eqnarray}
 \Ogw(f) = \frac{f}{\rho_c} \frac{d\rho_\text{GW}}{df}\,,
\end{eqnarray}
which represents the fractional contribution of gravitational waves to the critical energy density of the Universe \cite{Allen_Romano_1999}.
Here $d\rho_\text{GW}$ is the energy density in the frequency interval $f$ to $f+df$,  
$\rho _{c} = 3H_0^2c^2/(8\pi G)$ is the critical energy density of the Universe, and the Hubble parameter $H_0=67.9 \pm 0.55$ km s$^{-1}$ Mpc$^{-1}$ is taken from Planck~\cite{Planck_2015}.

In order to model the background from all the binary mergers in the Universe, we follow a similar approach to~\cite{gw150914_stoch}. 
A population of binary merger events may be characterized by a set of average source parameters $\theta$ (such as component masses or spins).
If such a population merges at a rate $R_m(z;\theta)$ per unit comoving volume per unit source time at a given redshift $z$, then the total gravitational-wave energy density spectrum from all the sources is given by (see, e.g.~\cite{Regimbau_2011,gw150914_stoch})
\begin{equation} \label{eq:Ogw}
\Ogw(f,\theta)=\frac{f}{\rho_c H_0} \int_0^{z_{\rm max}} dz \frac{R_m(z;\theta) dE_{\rm GW}(f_s;\theta)/df_s}{(1+z) E(\Omega_M,\Omega_\Lambda,z)}.  
\end{equation}
Here $dE_{\rm GW}(f_s,\theta)/df_s$ is the energy spectrum emitted by a single source evaluated in terms of the source frequency $f_s=(1+z)f$. 
The function $E(\Omega_M,\Omega_\Lambda,z)=\sqrt{\Omega_M(1+z)^3+\Omega_\Lambda}$ accounts for the dependence of comoving volume on redshift assuming the best-fit cosmology from Planck~\cite{Planck_2015}, where $\Omega_M=1-\Omega_\Lambda=0.3065$. 
We choose to cut off the redshift integral at $z_{\rm max}=10$. 
Redshifts larger than $z=5$ contribute little to the integral 
because of the $[(1+z)E(z)]^{-1}$ factor in Eq. \ref{eq:Ogw},
as well as the small number of stars formed at such high redshift, see for example \cite{Regimbau_2011, ZhuEA_2011, Rosado_2011, MarassiEA_2011, WuEA_2012, ZhuEA_2013, Kowolska_2015,CallisterEA_2016}. 

The energy spectrum $dE_{\rm GW}/df_s$ is determined from the strain waveform of the binary system. 
The dominant contribution to the background comes from the inspiral phase of the binary merger, for which $dE/df_s \propto \Mc^{5/3} f^{-1/3}$, where $\Mc=(m_1m_2)^{3/5}/(m_1+m_2)^{1/5}$ is the chirp mass for a binary system with component masses $m_1$ and $m_2$.
In the BNS case we only consider the inspiral phase, since neutron stars merge at $\sim 2$ kHz, well above the sensitive band of stochastic searches. We introduce a frequency cutoff at the innermost stable circular orbit.
For BBH events,
we include the merger and ringdown phases using the waveforms from \cite{AjithEA_2008, ZhuEA_2011} with the modifications from \cite{AjithEA_2011}. 

The merger rate $R_m(z;\theta)$ is given by 
\begin{equation}
R_m(z;\theta) = \int_{t_{\min}}^{t_{\max}} R_f(z_f;\theta)p(t_d;\theta) dt_d,
\end{equation}
where $t_d$ is the time delay between formation and merger of a binary, $p(t_d;\theta)$ is the time delay distribution given parameters $\theta$, $z_f$ is the redshift at the formation time $t_f=t(z)-t_d$, and $t(z)$ is the age of the Universe at merger. 
We assume that the binary formation rate $R_f(z_f;\theta)$ scales with the star formation rate. For the BNS background, we make similar assumptions to those used in \cite{gw150914_stoch}, which are outlined in what follows below. 
We adopt the star formation model of~\cite{VangioniEA_2015}, 
which produces very similar results as compared to the model described by~\cite{Madau_2014}.
We assume a time delay distribution $p(t_d) \propto 1/t_d$, for $t_{\min} < t_d < t_{\max}$. Here $t_{\min}$ is the minimum 
delay time between the binary formation and merger. 
We assume $t_{\rm min}=20\ {\rm Myr}$~\cite{MeacherEA_2015}. The maximum time delay $t_{\max}$ is set to the Hubble time~\cite{BelczynskiEA_2002,Ando_2004,BelczynskiEA_2006,deFreitasEA_2006,BergerEA_2007,Nakar_2007,O'ShaughnessyEA_2008,DominikEA_2012,DominikEA_2013}. 
We also need to consider the distribution of the component masses to calculate $\Ogw$.
We assume that each mass is drawn from uniform distribution ranging from 1 to 2 $M_\odot$.
The value of $R_m$ at $z=0$ is normalized to the median BNS merger rate implied by GW170817, which is $\bnsrate$~\cite{gw170817}.
This rate is slightly higher than the realistic BNS merger rate predictions of~\cite{rates2010}, and those adopted in previous studies (e.g. \cite{Regimbau_2011,Rosado_2011}), but is consistent with optimistic predictions. 

The calculation of the BBH background is similar, with the following differences. We assume $t_{\rm min}=50\ {\rm Myr}$ for the minimum time delay~\cite{DominikEA_2013,gw150914_stoch}. 
Massive black holes are formed preferentially in low-metallicity environments. For binary systems with at lease one black hole more massive than $30 M_\odot$, we therefore re-weight the star formation rate $R_f(z)$ by the fraction of stars with metallicities $Z\leq Z_\odot/2$. 
Following \cite{gw150914_stoch}, we adopt the mean metallicity-redshift relation of \cite{Madau_2014}, with appropriate scalings to account for local observations~\cite{BelczynskiEA_2016,VangioniEA_2015}.
For a consistent computation, a mass distribution needs to be specified.
We use a power-law distribution of the primary (i.e., larger mass) component $p(m_1) \propto m_1^{-2.35}$ and a uniform distribution of the secondary~\cite{gw170104,o1bbh}. 
In addition, we require that the component masses take values in the range $5-95\, M_\odot$ with $m_1+m_2<100\, M_\odot$ and $m_2 < m_1$, in agreement with the observations of BBHs to date~\cite{gw170104}. For the rate of BBH mergers, we use the most recent published result associated with the power-law mass distribution $\bbhrate$~\cite{gw170104,bbh}.
As shown in \cite{stoch_O1}, using a flat-log mass distribution instead of the power-law only affects $\Ogw(f)$ at frequencies above 100 Hz, which has very little impact on the detectability of the stochastic background with LIGO and Virgo. Frequencies below 100 Hz contribute to more than 99\% of the sensitivity of the stochastic search~\cite{stoch_O1}. 

The choices affecting the mass distribution have a minimal effect on the background energy density $\Omega_{\rm GW}$ in the low frequency part of the spectrum. 
We estimate the magnitude of the effect on the rate $R$ using the approximate scaling relationship $R\sim \langle VT \rangle^{-1} \sim \langle M_c^{5/2} \rangle ^{-1}$, where $VT$ is the sensitive spacetime volume of the instrument~\cite{Fishbach:2017zga, scaling}. 
In the BBH case for example, imposing a mass cutoff of $m_1,m_2<50~M_\odot$ to the power-law mass distribution reduces the estimate by less than 15\%. Similarly, using the same scaling law, we estimate that replacing a uniform mass distribution for BNS with a Gaussian mass distribution centered around $1.4 M_\odot$ leads to an increase of approximately 10\%. Both adjustments are well within the dominant statistical Poisson uncertainty.

{\em Predictions and detectability} ---
A stochastic background of gravitational-waves introduces a correlated signal in networks of terrestrial detectors.
This signal, expected to be much weaker than the
detector noise, can be distinguished from noise by cross-correlating the strain data from two or more detectors. 
For a network of $n$ detectors, 
assuming an isotropic, unpolarized, Gaussian, and stationary background, 
the optimal signal-to-noise ratio (SNR) of a cross-correlation search is given by
\begin{equation}
  \text{SNR} =\frac{3 H_0^2}{10 \pi^2} \sqrt{2T} \left[
\int_0^\infty df\>
\sum_{i=1}^n\sum_{j>i}
\frac{\gamma_{ij}^2(f)\Ogw^2(f)}{f^6 P_i(f)P_j(f)} \right]^{1/2}\,,
\label{eq:snrCC}
\end{equation}
in which $i,j$ run over detector pairs, $P_i(f)$ and $P_j(f)$ are the one-sided strain noise power spectral densities of the two detectors, and $\gamma_{ij}(f)$ is the normalized isotropic overlap reduction function between the pair~\cite{Allen_Romano_1999, gw150914_stoch}.
While the cross correlation search is not optimal for non-Gaussian backgrounds, Eq.~\ref{eq:snrCC} gives the correct expression for the cross-correlation signal-to-noise ratio irrespective of the Gaussianity of the background~\cite{MeacherEA_2014,MeacherEA_2015}.

On the left hand panel of Fig.~\ref{fig:sens}, we show the estimates on the background energy density $\Ogw(f)$ for the BNS and BBH merger populations described in the previous section (red and green curves, respectively). 
The total (combined) background from BBH and BNS mergers is also plotted (solid blue curve) along with the $90\%$ credible Poisson uncertainties in the local rate (indicated by the grey shaded region). Considering this uncertainty, we predict $\Ogw^{\text{tot}}(f=\unit[25]{Hz}) = \OmegaRange$.

The spectrum is well approximated by a power law $\Ogw (f) \propto f^{2/3}$ at low frequencies where the contribution from the inspiral phase is dominant. 
The power law remains a good approximation for $f \lesssim 100$ Hz.
Indeed, previous work suggests that the deviation from power-law behaviour is probably inaccessible by advanced detectors~\cite{CallisterEA_2016}.
The frequency region between 10 Hz and 100 Hz accounts for more than 99\% of the accumulated SNR for the Advanced LIGO-Virgo network~\cite{stoch_O1}.
The median value of the background from BBH mergers $[\Ogw^{\text{BBH}}(f=\unit[25]{Hz}) = \OmegaRangeBBH]$ is comparable with our previous predictions~\cite{gw150914_stoch} but the statistical uncertainty has decreased by a factor of 32\% due to an increase in the number of observed BBH mergers.
The median contribution from BNS systems is of the same order $[\Ogw{^\text{BNS}}(f=\unit[25]{Hz}) = \OmegaRangeBNS]$. The statistical uncertainty of the BNS background is larger than that of the BBH background, since we have only one BNS detection.  
These results are summarized in Table \ref{tab:results}.
 
\begin{table}
\caption{Estimates of the background energy density $\Ogw(f)$ at 25 Hz for each of the BNS, BBH and total background contributions, along with the 90\% Poisson error bounds. We also show the average time $\tau$ between events as seen by a detector in the frequency band above 10 Hz, and the number of overlapping sources at a given time $\lambda$. We quote the number given the median rate and associated Poisson error bounds. 
} \label{tab:results} 
\begin{tabular}{ccccccc}
 && $\Ogw(25\ {\rm Hz})$ && $\tau$ [s] && $\lambda$ \\ \hline
BNS		&& \OmegaRangeBNS && \tauBNS && \lambdaBNS \\
BBH		&& \OmegaRangeBBH && \tauBBH && \lambdaBBH \\
Total	&& \OmegaRange && \tauTotal  && \lambdaTot\\
\end{tabular}
\end{table}
 
The estimates on the background energy density shown in Fig.~\ref{fig:sens} are compared with  {\em power-law integrated} (PI) curves~\cite{locus} at various observing sensitivities (O2, O3 and Design).
The PI curves represent the expected $1\sigma$ sensitivity of the standard cross-correlation search to power-law gravitational-wave backgrounds~\cite{locus}, for example the 
$\Ogw(f) \propto f^{2/3}$ spectrum expected from the inspiral phase of binary mergers.
Hence, if a power-law spectrum intersects a PI curve, then it has ${\rm SNR}\ge 1$ for the corresponding observing run.

Although the stochastic background is dominated by unresolvable sources, the energy-density spectra in Fig. \ref{fig:sens} \textit{include} contributions from the loudest, individually detectable events. Simulations of the astrophysical background given the inferred rate and mass distribution indicate that removing sources that are individually detectable by the LIGO-Virgo network with a combined SNR $>12$ has a very small impact on the results. 
The detectable sources have an even smaller effect than shown by the analysis in~\cite{gw150914_stoch}, which considered only the population of loud, high mass sources.
This highlights the fact that the spectrum is dominated by low-mass systems, which are less likely to be detected individually.

Different assumptions are possible on the various distributions that enter the calculation of $\Ogw$ (Eq. \ref{eq:Ogw}), such as star formation rate, metallicity evolution and delay times.
However, we have verified that variations on our assumptions are contained within the Poisson band shown in Fig. \ref{fig:sens}, consistent with the detailed study in \cite{gw150914_stoch}.
Additionally, if some of the observed BBH systems were of primordial (rather than stellar) origin, with a different redshift distribution, their contribution to the stochastic background spectrum could be weaker~\cite{BirdEA_2016,SasakiEA_2016,MandicEA_2016}.

The right hand panel of Fig.~\ref{fig:sens} shows the expected accumulated SNR as a function of total observation time, updated from \cite{gw150914_stoch}. 
For O1, which lasted approximately 4 months, we use the actual instrumental sensitivities of the two LIGO detectors (Virgo was not yet operating at that time)\cite{stoch_O1}.
The second observing run (O2) was recently completed and ran for approximately 9 months. For this run we use typical sensitivities of 100 Mpc for Livingston and 60 Mpc for Hanford, assuming a duty cycle of 50\%. We do not include Virgo as it does not contribute significantly to the sensitivity of stochastic searches in O2 due to the lower range and one month integration time. For the next planned observing run (O3), and the following stages of sensitivity improvements, we include Virgo and assume a 50\% duty cycle for each detector. 
Following~\cite{prospects_2016}, we assume O3 will be 12 months long (2017 -- 2018). We define the ``Near Design'' phase (2019+) to be a 12 month run where Hanford and Livingston operate at design sensitivity and Virgo at late sensitivity. Lastly, we assume that the ``Design'' phase (2022+) will be 24 months and will have Hanford, Livingston, and Virgo operating at design sensitivity. These assumptions are broadly consistent with~\cite{prospects_2016}, though we make specific assumptions about the duration of each observing run for concreteness. 

The median total background from a combined BBH and BNS background may be identified with ${\rm SNR}=3$, corresponding to false alarm probability $<3\times 10^{-3}$, after approximately \obs\ of observing. 
In the most optimistic scenario allowed by statistical uncertainties, the total background could be identified after \obsopt\ or as early as O3. The most pessimistic case considered here is out of reach of the advanced detector network but is in the scope of third generation detectors~\cite{foreground}.

\begin{figure*}
\includegraphics[width=0.49\textwidth]{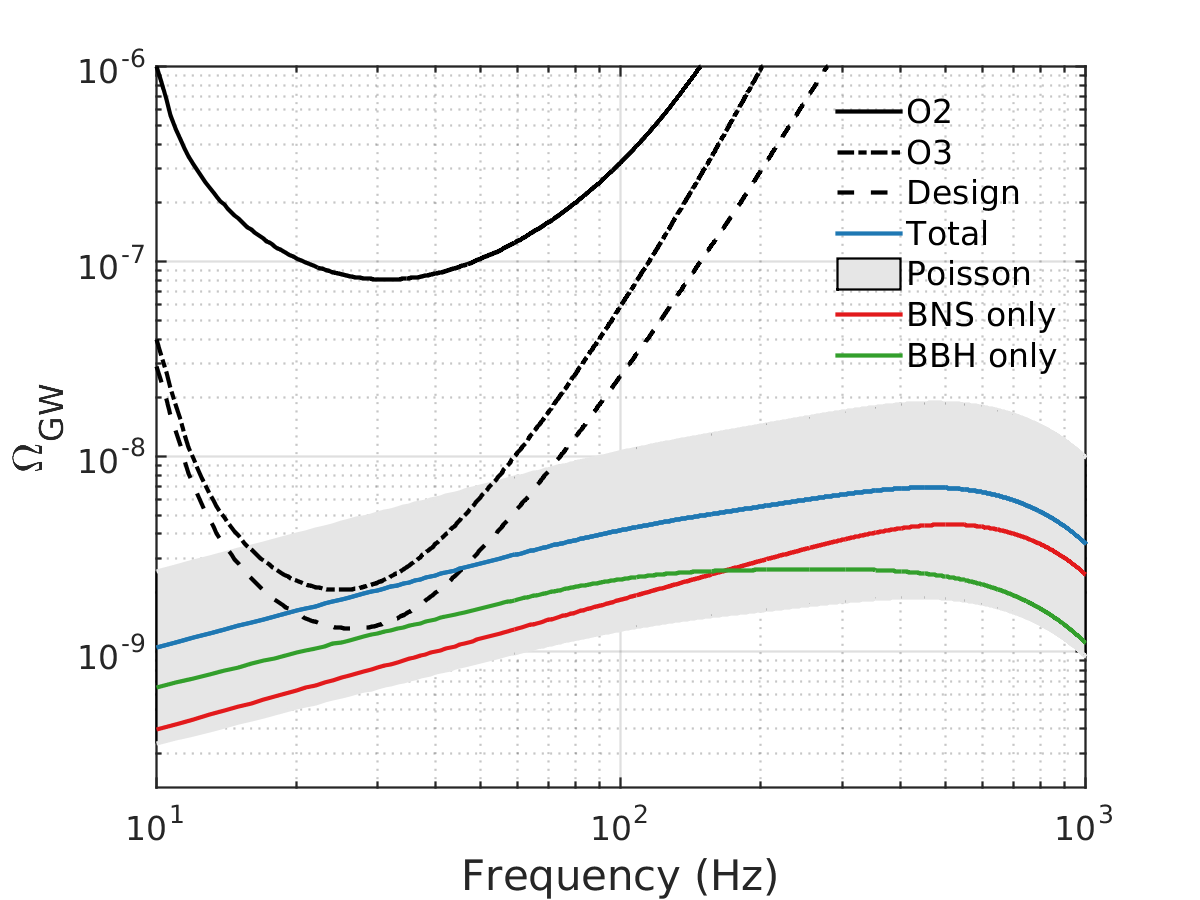}
\includegraphics[width=0.49\textwidth]{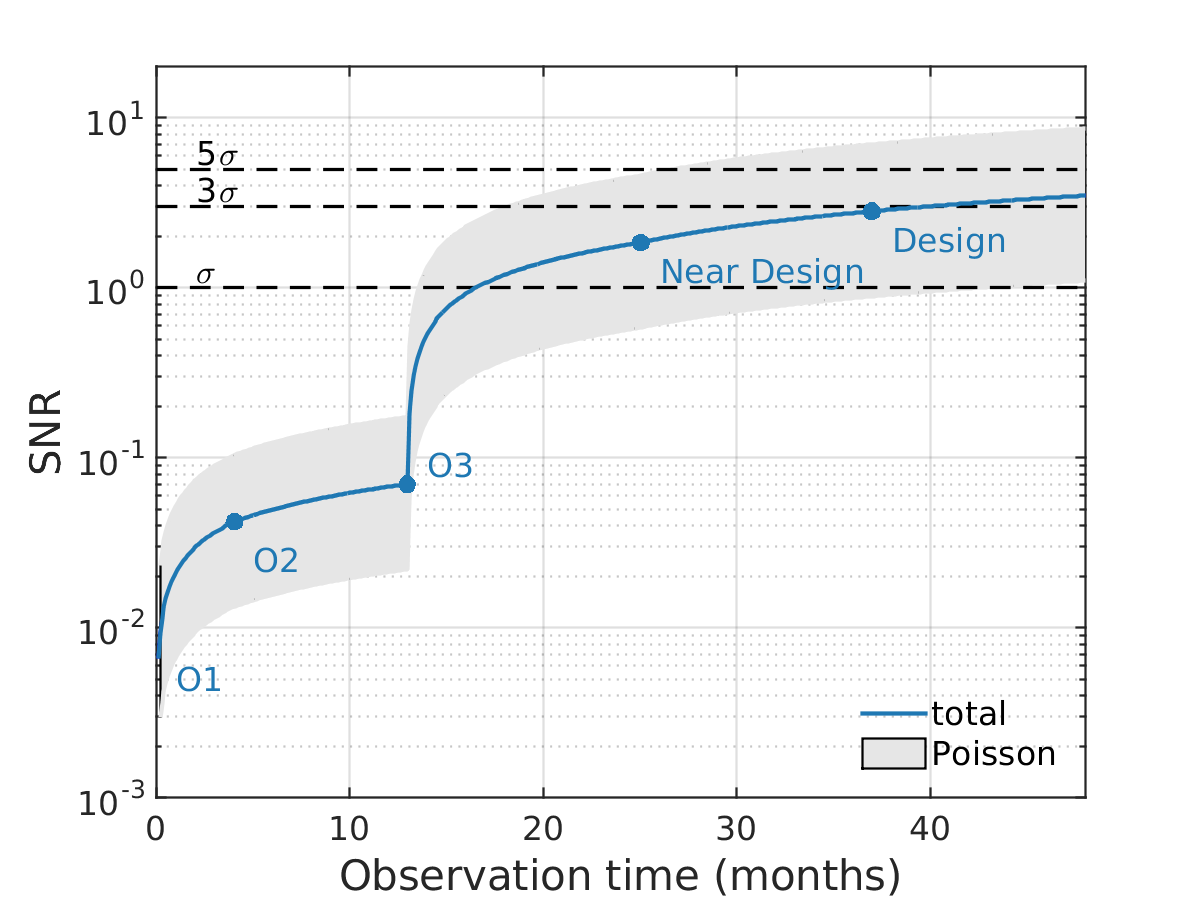}
\caption{The left panel shows the predicted median background for the BNS (red) and BBH (green) models described in the text, the total combined background (blue), and the Poisson error bars (grey shaded region) for the total background. We also show expected PI curves for observing runs O2, O3, and design sensitivity (see the main text for details about the assumptions made for these observing runs). Virgo is included in O3 and beyond. The PI curves for O3 and beyond cross the Poisson error region, indicating the possibility of detecting this background or placing interesting upper limits on the evolution of the binary merger rates with redshift. 
In the right panel, we plot the signal-to-noise ratio as a function of cumulative observing time for the median total background (blue curve) and associated uncertainty (shaded region). The median of the predicted total background can be detected with ${\rm SNR}=3$ after \obs\ of observation time, with LIGO-Virgo reaching design sensitivity (2022 -- 2024). The markers indicate the transition between observing runs. We only show 12 months of the Design phase for clarity, however this phase is 24 months long for the calculation of the PI curves (see \cite{prospects_2016}).} \label{fig:sens}
\end{figure*}

Although the BNS and BBH backgrounds have similar energy-densities, they have extremely different statistical properties. To illustrate this, we plot a simulated strain time-series in Fig.~\ref{fig:time_series} and show an example BNS (red) and BBH (green) background. 
The BNS events create an approximately continuous background consisting of a superposition of overlapping sources, since the duration of the waveform (\durBNS\ s on average in the frequency band above 10 Hz) is long compared to the average time interval between two successive events (\tauBNS\ s on average). Considering the uncertainty on the rate, the average number of overlapping sources can vary between 3 and 45. In Fig. \ref{fig:time_series}, 14 sources overlap on average. 

The BBH background is different in nature even though the resulting energy density spectrum is similar. BBH events create a highly non-stationary and non-Gaussian background (sometimes referred to as a ‘popcorn background’ in the literature), i.e. individual events are well separated in time, on top of the continuous background from contributed BNS inspirals. The duration of the waveform is much smaller for these massive sources (\durBBH\ s on average in the band above 10 Hz, considering both the power law mass distribution and the distribution in redshift~\cite{foreground}) and much less than the time interval between events (\tauBBH\ s on average) resulting in rare overlaps. 

Table \ref{tab:results} shows the estimated energy density at 25 Hz for each of the BNS, BBH and Total backgrounds. We also show the average time between events $\tau$ for each of these backgrounds as well as the average number of overlapping sources at any time $\lambda$, and the associated Poisson error bounds. The inverse of $\tau$ gives the rate of events in the Universe in s$^{-1}$. 

\begin{figure}
\includegraphics[width=0.49\textwidth]{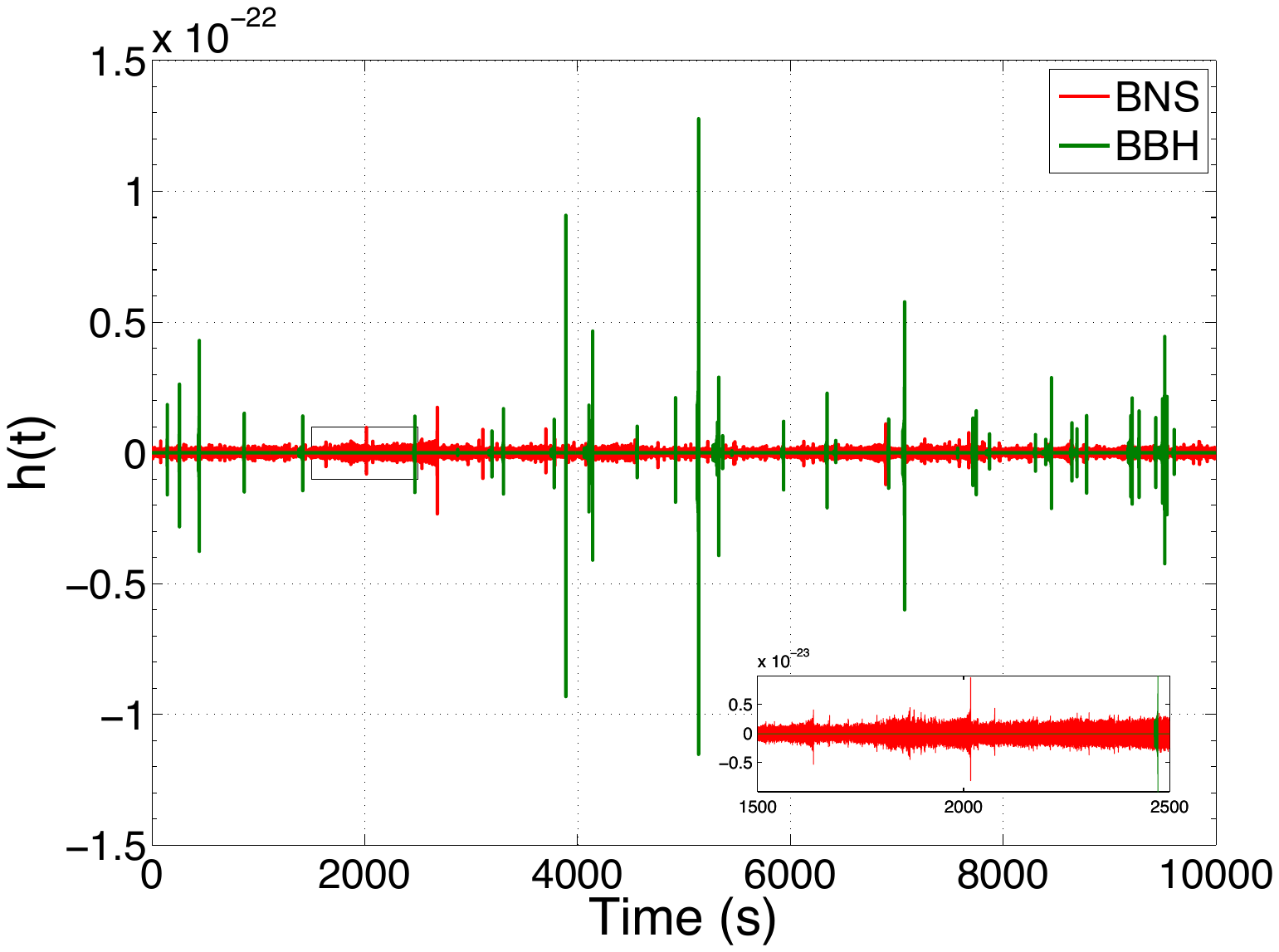}
\caption{We present a simulated time series of duration $10^4$ seconds illustrating the character of the BBH and BNS signals in the time domain. In red we show a simulated BNS background corresponding to the median rate as shown in Figure~\ref{fig:sens}, and in green we display the median BBH background. 
We do not show any detector noise, and do not remove some loud and close events that would be detected individually. The region in the black box, from 1800 -- 2600 seconds, is shown in greater detail in the inset. The BNS time series is continuous as it consists of a superposition of overlapping signals. On the other hand the BBH background (in green) is popcorn-like, and the signals do not overlap. Remarkably, even though the backgrounds have very different structure in the time domain, the energy in both backgrounds are comparable below 100\ {\rm Hz}, as seen in Figure~\ref{fig:sens}.}\label{fig:time_series}
\end{figure}

{\em Conclusion} ---
The first gravitational wave detection of a binary neutron star system implies a significant contribution to the stochastic gravitational wave background from BNS mergers. 
Assuming the median merger rates, the background may be detected with ${\rm SNR}=3$ after \obs\ of accumulated observation time, during the Design phase (2022+)\cite{prospects_2016}.
In the most optimistic case, an astrophysical background may be observed at a level of $3\sigma$ after only \obsopt\ of observation, during O3, the next observing run.

There are additional factors which may lead to an even earlier detection. First, the presence of additional sources, for example black hole-neutron star (BHNS) systems, will further add to the total background. Even small contributions to the background can decrease the time to detection. 
Previous estimates in the literature, for example \cite{o1rates,Kowalska:2012bb}, suggest that the BHNS will contribute roughly a few percent to the total background, although the uncertainty is large.
Second, the analysis we have presented here assumes the standard cross-correlation search. 
Specialized non-Gaussian searches may be more sensitive, particularly to the BBH background~\cite{popcorn,popcorn2,Smith:2017vfk}. 
Unlike a standard matched filter search, non-Gaussian pipelines do not attempt to find individual events, but rather to measure the rate of sub-threshold events independently of their distribution.

A detection of the astrophysical background allows for a rich set of follow-up studies to fully understand its composition. The difference in the time-domain structure of the BBH and BNS signals may allow the BNS and BBH backgrounds to be measured independently. After detecting the background, stochastic analyses can address whether the background is isotropic~\cite{sph_stoch,rad_stoch_1,rad_stoch_2}, unpolarized~\cite{axion_inflation}, and consistent with general relativity~\cite{TestingGR_stoch}. Finally, understanding the astrophysical background is crucial to subtract it and enable searches for a background of cosmological origin~\cite{foreground}.

\begin{acknowledgments}
\bigskip\noindent\textit{Acknowledgments} ---
The authors gratefully acknowledge the support of the United States
National Science Foundation (NSF) for the construction and operation of the
LIGO Laboratory and Advanced LIGO as well as the Science and Technology Facilities Council (STFC) of the
United Kingdom, the Max-Planck-Society (MPS), and the State of
Niedersachsen/Germany for support of the construction of Advanced LIGO 
and construction and operation of the GEO600 detector. 
Additional support for Advanced LIGO was provided by the Australian Research Council.
The authors gratefully acknowledge the Italian Istituto Nazionale di Fisica Nucleare (INFN),  
the French Centre National de la Recherche Scientifique (CNRS) and
the Foundation for Fundamental Research on Matter supported by the Netherlands Organisation for Scientific Research, 
for the construction and operation of the Virgo detector
and the creation and support  of the EGO consortium. 
The authors also gratefully acknowledge research support from these agencies as well as by 
the Council of Scientific and Industrial Research of India, 
the Department of Science and Technology, India,
the Science \& Engineering Research Board (SERB), India,
the Ministry of Human Resource Development, India,
the Spanish  Agencia Estatal de Investigaci\'on,
the  Vicepresid\`encia i Conselleria d'Innovaci\'o, Recerca i Turisme and the Conselleria d'Educaci\'o i Universitat del Govern de les Illes Balears,
the Conselleria d'Educaci\'o, Investigaci\'o, Cultura i Esport de la Generalitat Valenciana,
the National Science Centre of Poland,
the Swiss National Science Foundation (SNSF),
the Russian Foundation for Basic Research, 
the Russian Science Foundation,
the European Commission,
the European Regional Development Funds (ERDF),
the Royal Society, 
the Scottish Funding Council, 
the Scottish Universities Physics Alliance, 
the Hungarian Scientific Research Fund (OTKA),
the Lyon Institute of Origins (LIO),
the National Research, Development and Innovation Office Hungary (NKFI), 
the National Research Foundation of Korea,
Industry Canada and the Province of Ontario through the Ministry of Economic Development and Innovation, 
the Natural Science and Engineering Research Council Canada,
the Canadian Institute for Advanced Research,
the Brazilian Ministry of Science, Technology, Innovations, and Communications,
the International Center for Theoretical Physics South American Institute for Fundamental Research (ICTP-SAIFR), 
the Research Grants Council of Hong Kong,
the National Natural Science Foundation of China (NSFC),
the Leverhulme Trust, 
the Research Corporation, 
the Ministry of Science and Technology (MOST), Taiwan
and
the Kavli Foundation.
The authors gratefully acknowledge the support of the NSF, STFC, MPS, INFN, CNRS and the
State of Niedersachsen/Germany for provision of computational resources.

This  article  has  been  assigned the document number LIGO-P1700272.

\end{acknowledgments}

\bibliographystyle{apsrev4-1} 
\bibliography{implications}

\begin{thebibliography}{59}
\bibitem [{bbh()}]{bbh}%
  \BibitemOpen
  \href@noop {} {}\bibinfo {note} {An analysis using all events in O2 will be
  done in a follow-up paper. However, as noted in \cite{stoch_O1}, the
  power-law mass distribution considered here is dominated by events with low
  masses, and so the inclusion of GW170814 would not change the rate estimate
  significantly.}\BibitemShut {Stop}%
\bibitem [{sca()}]{scaling}%
  \BibitemOpen
  \href@noop {} {}\bibinfo {note} {In fact this scaling relationship is
  approximate, but leads to a conservative estimate.}\BibitemShut {Stop}%
\end{thebibliography}

\iftoggle{endauthorlist}{
  %
  %
  \let\author\myauthor
  \let\affiliation\myaffiliation
  \let\maketitle\mymaketitle
  \title{The LIGO Scientific Collaboration and Virgo Collaboration}
  \pacs{}




\author{%
B.~P.~Abbott,$^{1}$  
R.~Abbott,$^{1}$  
T.~D.~Abbott,$^{2}$  
F.~Acernese,$^{3,4}$ 
K.~Ackley,$^{5,6}$  
C.~Adams,$^{7}$  
T.~Adams,$^{8}$ 
P.~Addesso,$^{9}$  
R.~X.~Adhikari,$^{1}$  
V.~B.~Adya,$^{10}$
% ...
\\
\medskip
(LIGO Scientific Collaboration and Virgo Collaboration) 
\\
\medskip
{${}^{\dag}$Deceased, February 2017. }%
{${}^{\ddag}$Deceased, December 2016. }%
}\noaffiliation
\affiliation {LIGO, California Institute of Technology, Pasadena, CA 91125, USA }
\affiliation {Louisiana State University, Baton Rouge, LA 70803, USA }
\affiliation {Universit\`a di Salerno, Fisciano, I-84084 Salerno, Italy }
\affiliation {INFN, Sezione di Napoli, Complesso Universitario di Monte S.Angelo, I-80126 Napoli, Italy }
\affiliation {University of Florida, Gainesville, FL 32611, USA }
\affiliation {OzGrav, School of Physics \& Astronomy, Monash University, Clayton 3800, Victoria, Australia }
\affiliation {LIGO Livingston Observatory, Livingston, LA 70754, USA }
\affiliation {Laboratoire d'Annecy-le-Vieux de Physique des Particules (LAPP), Universit\'e Savoie Mont Blanc, CNRS/IN2P3, F-74941 Annecy, France }
\affiliation {University of Sannio at Benevento, I-82100 Benevento, Italy and INFN, Sezione di Napoli, I-80100 Napoli, Italy }
\affiliation {Max Planck Institute for Gravitational Physics (Albert Einstein Institute), D-30167 Hannover, Germany }
\affiliation {The University of Mississippi, University, MS 38677, USA }
\affiliation {NCSA, University of Illinois at Urbana-Champaign, Urbana, IL 61801, USA }
\affiliation {University of Cambridge, Cambridge CB2 1TN, United Kingdom }
\affiliation {Nikhef, Science Park, 1098 XG Amsterdam, The Netherlands }
\affiliation {LIGO, Massachusetts Institute of Technology, Cambridge, MA 02139, USA }
\affiliation {Instituto Nacional de Pesquisas Espaciais, 12227-010 S\~{a}o Jos\'{e} dos Campos, S\~{a}o Paulo, Brazil }
\affiliation {Gran Sasso Science Institute (GSSI), I-67100 L'Aquila, Italy }
\affiliation {INFN, Laboratori Nazionali del Gran Sasso, I-67100 Assergi, Italy }
\affiliation {Inter-University Centre for Astronomy and Astrophysics, Pune 411007, India }
\affiliation {International Centre for Theoretical Sciences, Tata Institute of Fundamental Research, Bengaluru 560089, India }
\affiliation {University of Wisconsin-Milwaukee, Milwaukee, WI 53201, USA }
\affiliation {Leibniz Universit\"at Hannover, D-30167 Hannover, Germany }
\affiliation {Universit\`a di Pisa, I-56127 Pisa, Italy }
\affiliation {INFN, Sezione di Pisa, I-56127 Pisa, Italy }
\affiliation {OzGrav, Australian National University, Canberra, Australian Capital Territory 0200, Australia }
\affiliation {Laboratoire des Mat\'eriaux Avanc\'es (LMA), CNRS/IN2P3, F-69622 Villeurbanne, France }
\affiliation {SUPA, University of the West of Scotland, Paisley PA1 2BE, United Kingdom }
\affiliation {LAL, Univ. Paris-Sud, CNRS/IN2P3, Universit\'e Paris-Saclay, F-91898 Orsay, France }
\affiliation {California State University Fullerton, Fullerton, CA 92831, USA }
\affiliation {European Gravitational Observatory (EGO), I-56021 Cascina, Pisa, Italy }
\affiliation {Chennai Mathematical Institute, Chennai 603103, India }
\affiliation {Universit\`a di Roma Tor Vergata, I-00133 Roma, Italy }
\affiliation {INFN, Sezione di Roma Tor Vergata, I-00133 Roma, Italy }
\affiliation {Universit\"at Hamburg, D-22761 Hamburg, Germany }
\affiliation {INFN, Sezione di Roma, I-00185 Roma, Italy }
\affiliation {Cardiff University, Cardiff CF24 3AA, United Kingdom }
\affiliation {Embry-Riddle Aeronautical University, Prescott, AZ 86301, USA }
\affiliation {Max Planck Institute for Gravitational Physics (Albert Einstein Institute), D-14476 Potsdam-Golm, Germany }
\affiliation {APC, AstroParticule et Cosmologie, Universit\'e Paris Diderot, CNRS/IN2P3, CEA/Irfu, Observatoire de Paris, Sorbonne Paris Cit\'e, F-75205 Paris Cedex 13, France }
\affiliation {Korea Institute of Science and Technology Information, Daejeon 34141, Korea }
\affiliation {West Virginia University, Morgantown, WV 26506, USA }
\affiliation {Universit\`a di Perugia, I-06123 Perugia, Italy }
\affiliation {INFN, Sezione di Perugia, I-06123 Perugia, Italy }
\affiliation {Syracuse University, Syracuse, NY 13244, USA }
\affiliation {University of Minnesota, Minneapolis, MN 55455, USA }
\affiliation {SUPA, University of Glasgow, Glasgow G12 8QQ, United Kingdom }
\affiliation {LIGO Hanford Observatory, Richland, WA 99352, USA }
\affiliation {Caltech CaRT, Pasadena, CA 91125, USA }
\affiliation {Wigner RCP, RMKI, H-1121 Budapest, Konkoly Thege Mikl\'os \'ut 29-33, Hungary }
\affiliation {Columbia University, New York, NY 10027, USA }
\affiliation {Stanford University, Stanford, CA 94305, USA }
\affiliation {Universit\`a di Camerino, Dipartimento di Fisica, I-62032 Camerino, Italy }
\affiliation {Universit\`a di Padova, Dipartimento di Fisica e Astronomia, I-35131 Padova, Italy }
\affiliation {INFN, Sezione di Padova, I-35131 Padova, Italy }
\affiliation {Institute of Physics, E\"otv\"os University, P\'azm\'any P. s. 1/A, Budapest 1117, Hungary }
\affiliation {Nicolaus Copernicus Astronomical Center, Polish Academy of Sciences, 00-716, Warsaw, Poland }
\affiliation {Rochester Institute of Technology, Rochester, NY 14623, USA }
\affiliation {University of Birmingham, Birmingham B15 2TT, United Kingdom }
\affiliation {INFN, Sezione di Genova, I-16146 Genova, Italy }
\affiliation {RRCAT, Indore MP 452013, India }
\affiliation {Faculty of Physics, Lomonosov Moscow State University, Moscow 119991, Russia }
\affiliation {SUPA, University of Strathclyde, Glasgow G1 1XQ, United Kingdom }
\affiliation {The Pennsylvania State University, University Park, PA 16802, USA }
\affiliation {OzGrav, University of Western Australia, Crawley, Western Australia 6009, Australia }
\affiliation {Department of Astrophysics/IMAPP, Radboud University Nijmegen, P.O. Box 9010, 6500 GL Nijmegen, The Netherlands }
\affiliation {Artemis, Universit\'e C\^ote d'Azur, Observatoire C\^ote d'Azur, CNRS, CS 34229, F-06304 Nice Cedex 4, France }
\affiliation {Institut FOTON, CNRS, Universit\'e de Rennes 1, F-35042 Rennes, France }
\affiliation {Washington State University, Pullman, WA 99164, USA }
\affiliation {University of Oregon, Eugene, OR 97403, USA }
\affiliation {Laboratoire Kastler Brossel, UPMC-Sorbonne Universit\'es, CNRS, ENS-PSL Research University, Coll\`ege de France, F-75005 Paris, France }
\affiliation {Carleton College, Northfield, MN 55057, USA }
\affiliation {OzGrav, University of Adelaide, Adelaide, South Australia 5005, Australia }
\affiliation {Astronomical Observatory Warsaw University, 00-478 Warsaw, Poland }
\affiliation {VU University Amsterdam, 1081 HV Amsterdam, The Netherlands }
\affiliation {University of Maryland, College Park, MD 20742, USA }
\affiliation {Center for Relativistic Astrophysics, Georgia Institute of Technology, Atlanta, GA 30332, USA }
\affiliation {Universit\'e Claude Bernard Lyon 1, F-69622 Villeurbanne, France }
\affiliation {Universit\`a di Napoli `Federico II,' Complesso Universitario di Monte S.Angelo, I-80126 Napoli, Italy }
\affiliation {NASA Goddard Space Flight Center, Greenbelt, MD 20771, USA }
\affiliation {Dipartimento di Fisica, Universit\`a degli Studi di Genova, I-16146 Genova, Italy }
\affiliation {RESCEU, University of Tokyo, Tokyo, 113-0033, Japan. }
\affiliation {Tsinghua University, Beijing 100084, China }
\affiliation {Texas Tech University, Lubbock, TX 79409, USA }
\affiliation {Kenyon College, Gambier, OH 43022, USA }
\affiliation {Departamento de Astronom\'{\i }a y Astrof\'{\i }sica, Universitat de Val\`encia, E-46100 Burjassot, Val\`encia, Spain }
\affiliation {Museo Storico della Fisica e Centro Studi e Ricerche Enrico Fermi, I-00184 Roma, Italy }
\affiliation {National Tsing Hua University, Hsinchu City, 30013 Taiwan, Republic of China }
\affiliation {Charles Sturt University, Wagga Wagga, New South Wales 2678, Australia }
\affiliation {Center for Interdisciplinary Exploration \& Research in Astrophysics (CIERA), Northwestern University, Evanston, IL 60208, USA }
\affiliation {University of Chicago, Chicago, IL 60637, USA }
\affiliation {Pusan National University, Busan 46241, Korea }
\affiliation {The Chinese University of Hong Kong, Shatin, NT, Hong Kong }
\affiliation {INAF, Osservatorio Astronomico di Padova, I-35122 Padova, Italy }
\affiliation {INFN, Trento Institute for Fundamental Physics and Applications, I-38123 Povo, Trento, Italy }
\affiliation {OzGrav, University of Melbourne, Parkville, Victoria 3010, Australia }
\affiliation {Universit\`a di Roma `La Sapienza,' I-00185 Roma, Italy }
\affiliation {Universit\'e Libre de Bruxelles, Brussels 1050, Belgium }
\affiliation {Sonoma State University, Rohnert Park, CA 94928, USA }
\affiliation {Departamento de Matem\'aticas, Universitat de Val\`encia, E-46100 Burjassot, Val\`encia, Spain }
\affiliation {Montana State University, Bozeman, MT 59717, USA }
\affiliation {Universitat de les Illes Balears, IAC3---IEEC, E-07122 Palma de Mallorca, Spain }
\affiliation {The University of Texas Rio Grande Valley, Brownsville, TX 78520, USA }
\affiliation {Bellevue College, Bellevue, WA 98007, USA }
\affiliation {Institute for Plasma Research, Bhat, Gandhinagar 382428, India }
\affiliation {The University of Sheffield, Sheffield S10 2TN, United Kingdom }
\affiliation {Dipartimento di Scienze Matematiche, Fisiche e Informatiche, Universit\`a di Parma, I-43124 Parma, Italy }
\affiliation {INFN, Sezione di Milano Bicocca, Gruppo Collegato di Parma, I-43124 Parma, Italy }
\affiliation {California State University, Los Angeles, 5151 State University Dr, Los Angeles, CA 90032, USA }
\affiliation {Universit\`a di Trento, Dipartimento di Fisica, I-38123 Povo, Trento, Italy }
\affiliation {Montclair State University, Montclair, NJ 07043, USA }
\affiliation {National Astronomical Observatory of Japan, 2-21-1 Osawa, Mitaka, Tokyo 181-8588, Japan }
\affiliation {Canadian Institute for Theoretical Astrophysics, University of Toronto, Toronto, Ontario M5S 3H8, Canada }
\affiliation {Observatori Astron\`omic, Universitat de Val\`encia, E-46980 Paterna, Val\`encia, Spain }
\affiliation {School of Mathematics, University of Edinburgh, Edinburgh EH9 3FD, United Kingdom }
\affiliation {University and Institute of Advanced Research, Koba Institutional Area, Gandhinagar Gujarat 382007, India }
\affiliation {IISER-TVM, CET Campus, Trivandrum Kerala 695016, India }
\affiliation {University of Szeged, D\'om t\'er 9, Szeged 6720, Hungary }
\affiliation {University of Michigan, Ann Arbor, MI 48109, USA }
\affiliation {Tata Institute of Fundamental Research, Mumbai 400005, India }
\affiliation {INAF, Osservatorio Astronomico di Capodimonte, I-80131, Napoli, Italy }
\affiliation {Universit\`a degli Studi di Urbino `Carlo Bo,' I-61029 Urbino, Italy }
\affiliation {INFN, Sezione di Firenze, I-50019 Sesto Fiorentino, Firenze, Italy }
\affiliation {Physik-Institut, University of Zurich, Winterthurerstrasse 190, 8057 Zurich, Switzerland }
\affiliation {American University, Washington, D.C. 20016, USA }
\affiliation {University of Bia{\l }ystok, 15-424 Bia{\l }ystok, Poland }
\affiliation {University of Southampton, Southampton SO17 1BJ, United Kingdom }
\affiliation {University of Washington Bothell, 18115 Campus Way NE, Bothell, WA 98011, USA }
\affiliation {Institute of Applied Physics, Nizhny Novgorod, 603950, Russia }
\affiliation {Korea Astronomy and Space Science Institute, Daejeon 34055, Korea }
\affiliation {Inje University Gimhae, South Gyeongsang 50834, Korea }
\affiliation {National Institute for Mathematical Sciences, Daejeon 34047, Korea }
\affiliation {NCBJ, 05-400 \'Swierk-Otwock, Poland }
\affiliation {Institute of Mathematics, Polish Academy of Sciences, 00656 Warsaw, Poland }
\affiliation {Hillsdale College, Hillsdale, MI 49242, USA }
\affiliation {Hanyang University, Seoul 04763, Korea }
\affiliation {Seoul National University, Seoul 08826, Korea }
\affiliation {NASA Marshall Space Flight Center, Huntsville, AL 35811, USA }
\affiliation {ESPCI, CNRS, F-75005 Paris, France }
\affiliation {Southern University and A\&M College, Baton Rouge, LA 70813, USA }
\affiliation {College of William and Mary, Williamsburg, VA 23187, USA }
\affiliation {Centre Scientifique de Monaco, 8 quai Antoine Ier, MC-98000, Monaco }
\affiliation {Indian Institute of Technology Madras, Chennai 600036, India }
\affiliation {IISER-Kolkata, Mohanpur, West Bengal 741252, India }
\affiliation {Whitman College, 345 Boyer Avenue, Walla Walla, WA 99362 USA }
\affiliation {Indian Institute of Technology Bombay, Powai, Mumbai, Maharashtra 400076, India }
\affiliation {Scuola Normale Superiore, Piazza dei Cavalieri 7, I-56126 Pisa, Italy }
\affiliation {Universit\'e de Lyon, F-69361 Lyon, France }
\affiliation {Hobart and William Smith Colleges, Geneva, NY 14456, USA }
\affiliation {OzGrav, Swinburne University of Technology, Hawthorn VIC 3122, Australia }
\affiliation {Janusz Gil Institute of Astronomy, University of Zielona G\'ora, 65-265 Zielona G\'ora, Poland }
\affiliation {University of Washington, Seattle, WA 98195, USA }
\affiliation {King's College London, University of London, London WC2R 2LS, United Kingdom }
\affiliation {Indian Institute of Technology, Gandhinagar Ahmedabad Gujarat 382424, India }
\affiliation {Indian Institute of Technology Hyderabad, Sangareddy, Khandi, Telangana 502285, India }
\affiliation {International Institute of Physics, Universidade Federal do Rio Grande do Norte, Natal RN 59078-970, Brazil }
\affiliation {Andrews University, Berrien Springs, MI 49104, USA }
\affiliation {Universit\`a di Siena, I-53100 Siena, Italy }
\affiliation {Trinity University, San Antonio, TX 78212, USA }
\affiliation {Van Swinderen Institute for Particle Physics and Gravity, University of Groningen, Nijenborgh 4, 9747 AG Groningen, The Netherlands }
\affiliation {Abilene Christian University, Abilene, TX 79699, USA }
\affiliation {Colorado State University, Fort Collins, CO 80523, USA }


  \newpage
  \maketitle
}

\end{document}